\documentclass[%
reprint,
amsmath,amssymb,
aps,
]{revtex4-2}

\usepackage{graphicx}% Include figure files
\usepackage{dcolumn}% Align table columns on decimal point
\usepackage{bm}% bold math

\usepackage{braket}
\usepackage{graphicx}
\usepackage{amsmath}
\usepackage{hyperref}
\usepackage{color}
\hypersetup{hypertex=true,
	colorlinks=true,
	linkcolor=blue,
	anchorcolor=blue,
	citecolor=blue,
	urlcolor=blue}
\begin{document}

\title{Continuously Expanding the Response Frequency of Rydberg Atom-Based
	Microwave Sensor by Using Quantum Mixer}%

\author{Sheng-Xian Xiao}
\affiliation{Chongqing Key Laboratory for Strongly Coupled Physics, Chongqing University, Chongqing, 401331, China}

\affiliation{Center of Modern Physics, Institute for Smart City of Chongqing University in Liyang, Liyang 213300, China}
\author{Tao Wang}
%\author{Tao Wang \CJKfamily{gbsn}(汪涛)}
\thanks{corresponding author: tauwaang@cqu.edu.cn}
\affiliation{Chongqing Key Laboratory for Strongly Coupled Physics, Chongqing University, Chongqing, 401331, China}

\affiliation{Center of Modern Physics, Institute for Smart City of Chongqing University in Liyang, Liyang 213300, China}
\begin{abstract}
Microwave electric (MW) field measurements utilizing Rydberg atoms have witnessed significant advancements, achieving remarkable sensitivity, albeit limited to discrete MW frequencies resonant with Rydberg states. Recently, various continuous-frequency measurement schemes have emerged. However, when the MW detuning surpasses 1 GHz, the sensitivity degrades by over an order of magnitude compared to resonant measurements. In this paper, we successfully extend the response frequency range by harnessing a controlled driving field in conjunction with a quantum mixer and heterodyne technology, theoretically enabling infinite scalability. Notably, second-order effects stemming from quantum mixing necessitate careful consideration to ensure accurate electric field measurements. In addition, compared to resonant measurements, the sensitivity decline for far-detuned MW fields exceeding 1 GHz is less than twice, representing a significant improvement of several orders of magnitude over alternative schemes. Furthermore, the sensitivity of far-detuned MW fields can be efficiently enhanced by augmenting the intensity and frequency of the controlled field. For detunings ranging from 100 MHz to 2 GHz, we present optimal sensitivity values and the corresponding methods to achieve them. Our findings pave the way for Rydberg atom-based MW receivers characterized by both high sensitivity and an exceptionally broad bandwidth.
\end{abstract}
\maketitle
%\tableofcontents

\section{INTRODUCTION}
Over the past decade, microwave (MW) field sensing based on Rydberg atoms has rapidly developed due to their exceptional sensitivity to electric fields \cite{1-Fan_2015,2-Yuan_2023,3-ZHANG20241515}. This technology leverages the four-level electromagnetically induced transparency (EIT) \cite{4-RevModPhys.77.633} and Autler–Townes (AT) splitting \cite{5-PhysRevA.81.053836} to realize MW sensors. Compared with traditional methods, Rydberg atom-based MW sensors offer advantages in repeatability, high sensitivity, self-calibration, and large bandwidth, making them promising candidates for the next generation of MW measurements \cite{3-ZHANG20241515}. They also exhibit broad application prospects in wireless communication \cite{6-10.1063/1.5031033,7-Song:19,8-9069423}, polarization measurements \cite{9-PhysRevLett.111.063001}, MW imaging \cite{10-Fan:14}and beyond.

However, achieving remarkable sensitivity is limited to discrete MW frequencies that are resonant or near-resonant with Rydberg states \cite{11-Sedlacek,12-Jing,13-10.1063/1.4947231,14-Kumar,15-10.1063/5.0069195,16-10.1063/5.0146768}. This constraint is inconsistent with the continuous frequencies required for many practical applications, thereby limiting the further utilization of Rydberg atomic microwave sensors. To address this issue, various schemes have been proposed, including the use of far-detuning AC Stark effects \cite{17-PhysRevApplied.15.014053,18-10.1063/5.0086357}, adjacent Rydberg resonance tuning \cite{19-PhysRevA.104.032824,20-PhysRevApplied.19.044049}, two-photon microwave transitions \cite{21-10.1063/1.4996234,22-PhysRevApplied.18.054003}, and the application of auxiliary microwave fields \cite{23-PhysRevA.103.063113,24-PhysRevA.107.043102}. Although these methods enable continuous frequency measurement, their sensitivity is generally reduced compared to resonance measurements, particularly for extended ranges beyond 1 GHz, with some cases observing a decrease in sensitivity by an order of magnitude. Notably, a scheme that achieves a large frequency extension range while maintaining high sensitivity over this range remains elusive.

In this paper, we propose an extended frequency scheme by incorporating quantum mixer \cite{25-PhysRevX.12.021061} with atomic heterodyne methods \cite{12-Jing}, facilitated by the application of an additional driving controlled field. The controlled field arises from the AC Stark shift induced by a low-frequency, weak radio-frequency (RF) field on the Rydberg state \cite{26-Bason_2010,27-Liu_2023,28-Miller_2016,29-PhysRevA.94.023832,30-10.1063/5.0162101}. Given the prevalence of quantum mixing in our approach, second-order effects stemming from the RF field cannot be overlooked, in contrast to their frequent neglect in other frameworks \cite{27-Liu_2023,30-10.1063/5.0162101}. By precisely adjusting the RF field, our scheme attains high sensitivity, a substantial extended frequency range, and maintains a sensitivity drop of less than a factor of three over an extended range of 2 GHz, representing at least an order of magnitude improvement over existing schemes.

The remainder of this article is organized as follows. In Sec.\ref{sec2}, we introduce our scheme in detail. In Sec.\ref{sec3},by using the theory of quantum frequency mixing, we derive the effective Hamiltonian. Base on this, the condition of controlled field is obtained. And we emphasize the significance of second-order terms arising from quantum mixing. Sec.\ref{sec4} delves into the relationship between the extended frequency range and sensitivity. Finally, in Sec.\ref{sec5}, we summarize our key findings and discuss potential avenues for future work.

\section{THE SCHEME \label{sec2}}

\begin{figure}
	\includegraphics[width=1\linewidth]{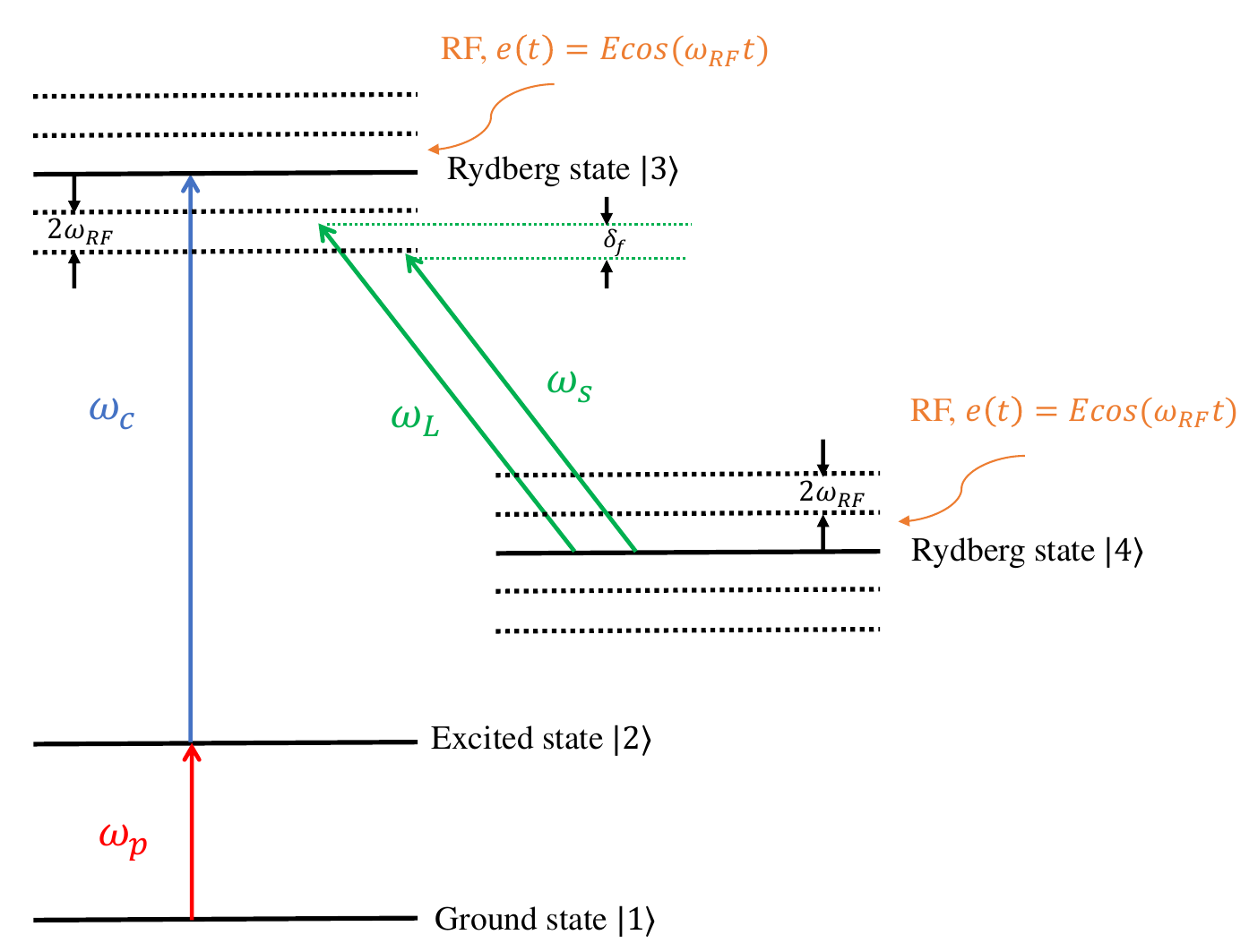}
	\caption{The scheme of Rydberg atom-based
		microwave sensor by combining quantum mixer and atomic heterodyne.}
	\label{fig1}
\end{figure}

Our proposed scheme is a typical four-level system as depicted in Fig.\ref{fig1}, in which the weak probe laser with frequency $\omega_p$ is scanned through resonance with the transition between the ground state $\left| 1\right\rangle $, and the excited state $\left| 2\right\rangle $, and the coupling laser with frequency $\omega_c$ is coupled the excited state $\left| 2\right\rangle $ and the Rydberg state $\left| 3\right\rangle $. Notably, unlike in the resonant case, the two Rydberg states $\left| 3\right\rangle $ and $\left| 4\right\rangle $ are dressed by a far-detuning MW electric field, which renders the measurement of MW electric field based on EIT-AT splitting ineffective. 

To restore the resonant response in the EIT-AT spectrum, we introduce an additional RF field, whose frequency and intensity are significantly smaller than the energy level difference utilized. This RF field can be regarded as a perturbation term, inducing a time-dependent energy shift in the atomic states due to the AC Stark effect. Given the negligible polarizability of the ground and excited states, their energy shifts can be disregarded. Consequently, we focus solely on the energy shifts of the two Rydberg states, which exhibit heightened sensitivity to electric fields. These shifts are given by  $-\alpha_{3}e^2(t)/2$ and $-\alpha_{4}e^2(t)/2$ \cite{27-Liu_2023,30-10.1063/5.0162101} , where $\alpha_{3(4)}$ is the polarizability of the Rydberg state $\left| 3\right\rangle(\left| 4\right\rangle)$ and $e(t)=E\cos(\omega_{RF}t)$ represents the electric component of the RF field. Therefore, the controlled Hamiltonian $H_C(t)$ can be written as:
\begin{eqnarray}
	H_C(t)=&\hbar A(1+\cos\omega t)\left| 3\right\rangle \left\langle 3\right|\nonumber
	\\&+\hbar A^{\prime}(1+\cos\omega t) \left| 4\right\rangle \left\langle 4\right|. \label{eq1}
\end{eqnarray}
The parameters $A=-\alpha_{3}E^2/4\hbar$, $A^{\prime}=-\alpha_{4}E^2/4\hbar$ and $\omega=2\omega_{RF}$ can be adjusted by RF field. $\hbar$ is reduced Planck's constant.
Then, in the interaction representation, we can write the far-detuning Hamiltonian of the system after rotating wave approximation (RWA) as :
\begin{align}
		H(t)=&-\frac{\hbar\Omega_p}{2}e^{i\Delta_pt}\left| 1\right\rangle \left\langle 2\right|+h.c.\nonumber
		\\&-\frac{\hbar\Omega_c}{2}e^{i\Delta_ct}\left| 2\right\rangle \left\langle 3\right|+h.c.\nonumber
		\\ &-\frac{\hbar\Omega_M}{2}e^{i\Delta_Mt}\left| 3\right\rangle \left\langle 4\right|+h.c.\nonumber
		\\ &+H_C(t),\label{eq2}
\end{align}
where $\Omega_p$, $\Omega_c$ and $\Omega_M$ are the Rabi frequencies corresponding to the transition $\left| 1\right\rangle \longleftrightarrow \left| 2\right\rangle$, $\left| 2\right\rangle \longleftrightarrow \left| 3\right\rangle$ and  $\left| 3\right\rangle \longleftrightarrow \left| 4\right\rangle$, respectively. The detunings of the probe laser, coupling laser, and MW electric field are denoted by $\Delta_p=\omega_p-\omega_{21}$, $\Delta_c=\omega_c-\omega_{32}$ and $\Delta_M=\omega_{34}-\omega_M$, respectively, where $\omega_{ij}=\omega_i-\omega_j$ is the energy level transition frequency. In this paper, we focus on the far-detuning MW, i.e., $\Omega_M\ll \Delta_M$.

In addition, to enhance the sensitivity of detecting the far-detuning MW electric field, we combine the quantum mixer and atomic heterodyne. As shown in Fig.\ref{fig1}, a strong local MW field is also applied. Consequently, $\Omega_M$ consists of two parts,namely, $\Omega_M=\Omega_L+\Omega_se^{i\delta_ft}$, where $\Omega_L\gg\Omega_s$ and $\delta_f$ is the frequency difference of the local field and signal field \cite{12-Jing}. Since $\delta_f$ is much smaller than the dynamic energy characteristic scale of the four-level EIT system, we can derive the EIT spectrum in the adiabatic limit.   

\section{The effective model \label{sec3}}

Since $H_C(t)$ doesn't commute with the third term of Eq.(\ref{eq2}), which describes the MW field, the mixing of these two high-frequency terms causes the resonant response of the system to reappear. This is the core of our method. According to the theory of quantum frequency mixing \cite{25-PhysRevX.12.021061}, we can expand the system's response frequency to the MW  field by precisely adjusting the controlled field, that is, the RF filed. In this section, by obtaining the effective Hamiltonian of our scheme, we can determine the parameters of the RF field that restore the resonant response in the EIT-AT spectrum. 

$A$ or $A^{\prime}$ is generally as large as or even much larger than $\omega$ in the laboratory. Therefore, to satisfy the high-frequency condition for using the theory of quantum frequency mixing, we need to perform a unitary transformation on Eq.(\ref{eq2}):
\begin{align}
	H_R(t)=&U^{\dagger}(t)\left( H(t)-i\hbar\frac{\partial}{\partial t}\right) U(t)\nonumber\\
	=&-\frac{\hbar\Omega_p}{2}e^{i\Delta_pt}\left| 1\right\rangle \left\langle 2\right|+h.c. \nonumber\\
	&-\sum_n\frac{\hbar\Omega_c\mathbf{J}_n(A/\omega)}{2}e^{i(\Delta_c-A-n\omega)t}\left| 2\right\rangle \left\langle 3\right|+h.c. \nonumber \\
	&-\sum_m\frac{\hbar\Omega_M\mathbf{J}_m(a/\omega)}{2}e^{i(\Delta_M-a-m\omega)t}\left| 3\right\rangle \left\langle 4\right|+h.c.\label{eq6}
\end{align}
where $a=A^{\prime}-A$ and $U(t)=\exp\left[ -\frac{i}{\hbar
}\int_{0}^{t}H_C(t) d\tau\right]$. In the deriving, we use the identity relation : $e^{iz\sin\omega t}=\sum_n \mathbf{J}_n(z)e^{in\omega t}$ with the Bessel functions of the first kind $\mathbf{J}_n(z)$.

The heterodyne detection requires fixed frequencies of probe laser and coupling laser, meaning that we should chose one of the Floquet side bands \cite{27-Liu_2023} to detect. In our scheme, we can set $\Delta_p$, $\Delta_c-A$ (i.e., $n=0$) to be low frequencies by adjusting the coupled laser, and we suppose RF field satisfies the condition $\Delta_M-a-k\omega\approx0$. Hence,
considering second-order effects and using the energy scale analysis of multi-mode Floquet theory to keep the low-frequency terms (see Appendix.\ref{ap1}), i.e., $\omega\gg\Omega_p,\Omega_c\mathbf{J}_n(A/\omega),\Omega_M\mathbf{J}_m(a/\omega)$, we can obtain the time-dependent effective Hamiltonian: 
\begin{align}
	H^{eff}_R(t)=&-\frac{\hbar\Omega_p}{2}e^{i\Delta_pt}\left| 1\right\rangle \left\langle 2\right|+h.c. \nonumber\\
	&-\frac{\hbar\Omega_c\mathbf{J}_0(A/\omega)}{2}e^{i(\Delta_c-A)t}\left| 2\right\rangle \left\langle 3\right|+h.c. \nonumber \\
	&-\frac{\hbar\Omega_M\mathbf{J}_k(a/\omega)}{2}e^{i(\Delta_M-a-k\omega)t}\left| 3\right\rangle \left\langle 4\right|+h.c.\nonumber\\
	&+H^{(2),R}_0\label{eq7}
\end{align}
The second-order term $H^{(2),R}_0$ comes from the mixing of the high-frequency terms, and is given by Eq.(\ref{eq5}):
\begin{align}
	H^{(2),R}_0=&-\sum_{n\neq0}\frac{\left[ \frac{\Omega_c\mathbf{J}_n(A/\omega)}{2}\left| 3\right\rangle \left\langle 2\right|,\frac{\Omega_c\mathbf{J}_n(A/\omega)}{2}\left| 2\right\rangle \left\langle 3\right|\right] }{\Delta_c-A-n\omega}\nonumber\\
	&-\sum_{m\neq k}\frac{\left[ \frac{\Omega_M\mathbf{J}_m(a/\omega)}{2}\left| 4\right\rangle \left\langle 3\right|,\frac{\Omega_M\mathbf{J}_m(a/\omega)}{2}\left| 3\right\rangle \left\langle 4\right|\right] }{\Delta_M-A^{\prime}+A-m\omega}\nonumber\\
	=&\sum_{n\neq0}\frac{\Omega_c^2\mathbf{J}_n^2(A/\omega)}{4(\Delta_c-A-n\omega)}(\left| 2\right\rangle \left\langle 2\right|-\left| 3\right\rangle \left\langle 3\right|)\nonumber\\
	&\sum_{m\neq k}\frac{\Omega_M^2\mathbf{J}_m^2(a/\omega)}{4(\Delta_M-A^{\prime}+A-m\omega)}(\left| 3\right\rangle \left\langle 3\right|-\left| 4\right\rangle \left\langle 4\right|)\label{A1}\\
	\approx&\sum_{m\neq k}\frac{\Omega_M^2\mathbf{J}_m^2(a/\omega)}{4(\Delta_M-A^{\prime}+A-m\omega)}(\left| 3\right\rangle \left\langle 3\right|-\left| 4\right\rangle \left\langle 4\right|)\nonumber\\
	=&\frac{\delta_M}{2}(\left| 3\right\rangle \left\langle 3\right|-\left| 4\right\rangle \left\langle 4\right|)\label{A2}
\end{align}
From Eq.(\ref{A1}) to Eq.(\ref{A2}), we use the low-frequency condition $\Delta_c-A\approx0$. This also explains why we chose $n=0$ in Eq.(\ref{eq7}), that is, to ignore the second-order term brought by the coupling laser.

Then, through the rotation transformation, we can obtain the time-independent effective Hamiltonian $H^{eff}$ corresponding to Eq.(\ref{eq7}):
\begin{equation}
	-\hbar\left( 
	\begin{array}{cccc}
		0 & \frac{\Omega_p}{2} & 0 & 0\\
		\frac{\Omega_p}{2} & \Delta_p & \frac{\mathbf{J}_0(A/\omega)\Omega_c}{2} & 0\\
		0 &  \frac{\mathbf{J}_0(A/\omega)\Omega_c}{2} & \Delta_p+\delta_c& \frac{\mathbf{J}_k(a/\omega)\Omega_M}{2}\\
		0 & 0 & \frac{\mathbf{J}_k(a/\omega)\Omega_M}{2} &\Delta_p+\delta_c+\Delta_M^{eff}
	\end{array}
	\right), \label{eq9}
\end{equation}
where $\delta_c=\Delta_c-A-\frac{\delta_M}{2}$ and $\Delta_M^{eff}=\Delta_M-a-k\omega+\delta_M$ are the effective detunings of coupling laser and MW field, respectively. 
\begin{figure}
	\centering
	\includegraphics[width=1\linewidth]{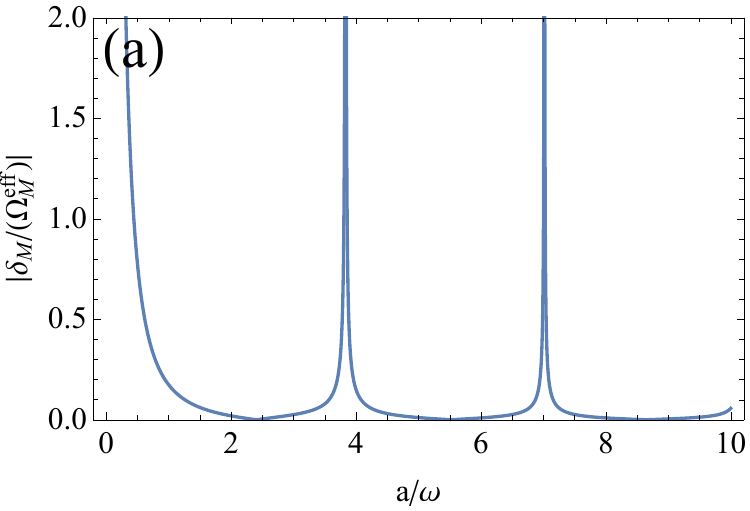}
	\includegraphics[width=1\linewidth]{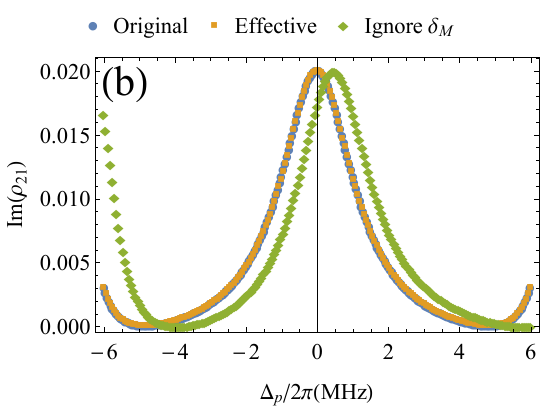}
	\includegraphics[width=1\linewidth]{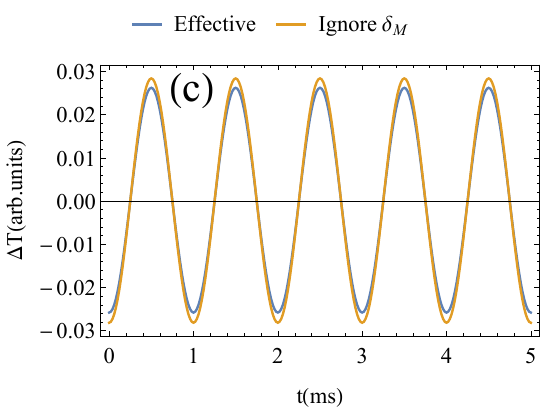}
	\caption{\label{fig2}(a) The Upper bound of $|\delta_M/(\mathbf{J}_k(a/\omega)\Omega_M)|$ changes with $a/\omega$ for $k=1$. (b) The imaginary part of $\rho_{21}$ changing with $\Delta_p$. System parameters: $\Omega_p=0.1\times2\pi$ MHz, $\Omega_c=10\times2\pi$ MHz, $\Delta_c=(5+1.8135/2)\times2\pi$ MHz, $\gamma_1=0$, $\gamma_2=5\times2\pi$ MHz, $\gamma_3=\gamma_4=0.003\times2\pi$ MHz. MW field parameters: $\Omega_M=40\times2\pi$ MHz, $\Delta_M=600\times2\pi$ MHz. RF field parameters: $\omega=401.209\times2\pi$ MHz for $k=1$, $A=5\times2\pi$ MHz, $a=0.5\omega$. $\delta_M=1.8135\times2\pi$ MHz. (c) The heterodyne EIT signal of the effective mode and the model ignoring $\delta_M$. $\Omega_s=1\times2\pi$ MHz, $\delta_f=1\times2\pi$ kHz, and the other parameters are same as those in (b). }
\end{figure} 

By precisely adjusting the coupling laser and the RF field, $\delta_c$ and $\Delta_M^{eff}$ can be set to 0. Therefore, the EIT spectrum of the system described by Eq.(\ref{eq9}) recovers the resonant response to MW field and the EIT-AT spectrum describes the effective Rabi frequency of MW signal field, $\mathbf{J}_k(a/\omega)\Omega_s$. This completely proves the reliability of our scheme. Based on this analysis, we can get the extended frequency of the resonant response: 
\begin{equation}
	\Delta_M = a+k\omega-\delta_M.\label{eq10}
\end{equation}
Namely, the extended frequency range is theoretically infinite due to the integer $k$. Furthermore, it also indicates that the control field, i.e., the RF field must strictly satisfy this relation.

From Eq.(\ref{eq10}), it can be seen that the second order term affects the parameter adjustment of the RF field. In previous studies \cite{27-Liu_2023,30-10.1063/5.0162101}, the second-order term was typically not considered because of their small value. However, it can't be ignored under certain parameter conditions in EIT-based heterodyne detection. To prove it, we calculate the ratio of $\delta_M$ and $\Omega_M\mathbf{J}_k(a/\omega)$: 
\begin{align}
	\left| \delta_M/(\mathbf{J}_k(a/\omega)\Omega_M)\right| \approx&\left| \sum_{m\neq k}\frac{\Omega_M\mathbf{J}_m^2(a/\omega)}{-2(m-k)\omega\mathbf{J}_k(a/\omega)}\right| \nonumber\\
	\le&\left| \sum_{m\neq k}\frac{0.1\mathbf{J}_m^2(a/\omega)}{-2(m-k)\mathbf{J}_k^2(a/\omega)}\right| \label{eq11}
\end{align}
In the above deriving, we regard $\Omega_M\mathbf{J}_k(a/\omega)\ll\omega$ as $\Omega_M\mathbf{J}_k(a/\omega)\le0.1\omega$. This sets an upper bound on the ratio of $\delta_M$ and $\Omega_M\mathbf{J}_k(a/\omega)$. 
%Whether the second-order term can be ignored depends on whether $\delta_M$ is much smaller than the effective Rabi frequency of MW field $\mathbf{J}_k(a/\omega)\Omega_M$. 
As shown in Fig.\ref{fig2} (a), the upper bound is not always much less than 1, especially for certain values of $a/\omega$, such as 0.1, 3.83, 7, $\dots$, where these upper bounds exceed 1. Since $\delta_M$ is one part of the effective detuning $\Delta_M$, ignoring $\delta_M $ will lead asymmetrical EIT-AT splitting \cite{19-PhysRevA.104.032824}.  These cases would be possible because the intensity and frequency of RF are not arbitrary and need to meet the conditions of weak field and low frequency with relative to the energy level difference. 

For further verification, using the master equation for the four-level density matrix $\rho$ \cite{31-10.1063/1.4984201}:
\begin{equation}
	\dot{\rho}=\frac{i}{\hbar}\left[ \rho,H\right] +\mathcal{D}(\rho),
\end{equation}
we numerically calculate the steady-state solutions of the original model, the effective model and the model ignoring the second-order terms. The matrix $\mathcal{D}(\rho)$ is given by:
\begin{equation}
	\left[\begin{array}{cccc}
		\gamma_{2} \rho_{22} & -\gamma_{12} \rho_{12} & -\gamma_{13} \rho_{13} & -\gamma_{14} \rho_{14} \\
		-\gamma_{21} \rho_{21} & \gamma_{3} \rho_{33}-\gamma_{2} \rho_{22} & -\gamma_{23} \rho_{23} & -\gamma_{24} \rho_{24} \\
		-\gamma_{31} \rho_{31} & -\gamma_{32} \rho_{32} & \gamma_{4} \rho_{44}-\gamma_{3} \rho_{33} & -\gamma_{34} \rho_{34} \\
		-\gamma_{41} \rho_{41} & -\gamma_{42} \rho_{42} & -\gamma_{43} \rho_{43} & -\gamma_{4} \rho_{44}
	\end{array}\right],
\end{equation} 
where $\gamma_{ij}=(\gamma_{i}+\gamma_{j})/2$ and $\gamma_{1,2,3,4}$ are the decay rate of the four levels. The imaginary part of $\rho_{21}$ varying with $\Delta_p$ is shown in Fig.\ref{fig2} (b). The numerical results clearly demonstrate that the results of the original Hamiltonian Eq.(\ref{eq2}) and the effective Hamiltonian Eq.(\ref{eq9}) agree very well, while the result of ignoring $\delta_M$ deviates from the first two. 

Since Im$[\rho_{21}]$ actually reflects the probe laser transmission $T$ and the heterodyne method aims to detect the changes of the  transmission $\Delta T$, ignoring the second-order term will lead to serious distortion of measurement veracity. To prove this, based on the parameters of the cesium atoms in a vapour cell at 
room temperature, we simulate heterodyne signal $\Delta T(t)$ in the adiabatic limit. As shown in  Fig.\ref{fig2} (c), for the same signal field strength $\Omega_s$, the amplitudes of $\Delta T(t)$  differ significantly between the effective model and the modle ignoring $\delta_M$. Therefore, the second-order term should be seriously considered in the scheme combining quantum mixer and atomic heterodyne for detecting far-detuning MW fields.

\section{Sensitivity and extended frequency range\label{sec4}} 

\begin{figure}
	\includegraphics[width=1\linewidth]{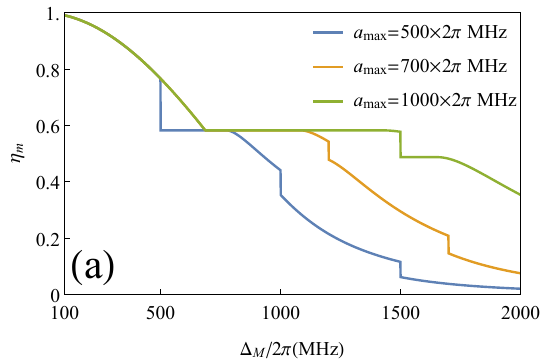}
	\includegraphics[width=1\linewidth]{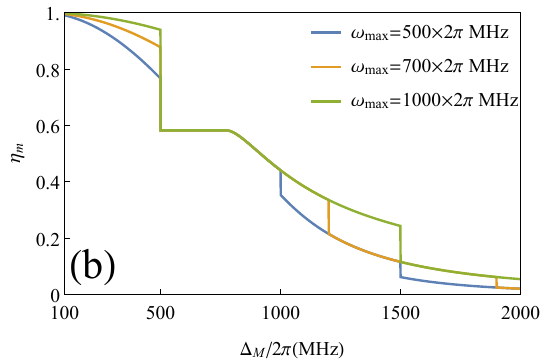}
	\caption{\label{fig3}The relationship between the modification coefficient of sensitivity $\eta_m$ and extended frequency range $\Delta_M$. (a) $\omega_{max}=500\times2\pi$ MHz. (b) $a_{max}=500\times2\pi$ MHz. $\omega_{min}=100\times2\pi$ MHz both in (a) and (b). }
	\label{fig3}
\end{figure}

Compared to the case of resonance, the sensitivity of our scheme for measuring the far-detuning MW field is somewhat weakened, as the effective Rabi frequency of MW field is modified by $\mathbf{J}_k(a/\omega)$ according to Eq.(\ref{eq9}). This is the necessary compromise for continuously extending the response frequency range and is also an inevitable consequence of quantum mixing technology. In this section, we present the correspondence between the extended frequency range and the optimal sensitivity that can be achieved. 

The sensitivity of our scheme can be obtained by dividing the resonance sensitivity by $\mathbf{J}_k(a/\omega)$, provided that systematic errors due to the introduction of RF fields are not considered. Therefore, within the allowable range of parameters, the best sensitivity can be achieved by identifying the RF field parameters that maximize $\mathbf{J}_k(a/\omega)$, that is, $a$ and $\omega$. In our scheme, the frequency and the intensity of RF field are very small compared to the energy level difference used, leading to practical limitations on the maximum values of $\omega$ and $a$ that can be achieved in the laboratory. Furthermore, due to the high-frequency requirements of quantum mixing theory,  there exists a minimum value for $\omega$.  Under such conditions, for a given MW detuning $\Delta_M$, the optimal modification coefficient $\eta_m$  is calculated as follows:
\begin{equation}
\eta_m=Max \left\lbrace \left| \mathbf{J}_k(a/\omega)\right|\right\rbrace.  
\end{equation}
By considering all $\omega$, $a$ and $k$ that satisfy the condition $0<a\le a_{max}$,  $\omega_{min}\le\omega\le\omega_{max}$ and $\Delta_M=a+k\omega$ to find the maximum of $|\mathbf{J}_k(a/\omega)|$ and corresponding $a$ and $\omega$, we can determine $\eta_m$ and the optimal RF filed parameters. Here, we ignore the $\delta_M$ because it is at least one order of magnitude smaller than $\omega$, resulting in a negligible impact on the modification coefficient $\mathbf{J}_k(a/\omega)$.

In Fig.\ref{fig3}, we exhibit the relationship between the sensitivity modification coefficient and the extended frequency range. As can be seen, when $\omega_{max}$ is fixed, with the increase of $a_{max}$, there is a significant enhancement in sensitivity, especially at the large detunings $\Delta_M$. Similarly, when $\omega_{max}$ is fixed, increasing $\omega_{max}$ also improves sensitivity. Therefore, increasing the intensity and frequency limit of the applied RF field will aid in realizing the MW field measurement with high sensitivity and highly wide bandwidth. In addition, it can be seen that the sensitivity changes abruptly at certain detuning points, primarily due to changes in the integer $k$. For example, for $a_{max}=500\times2\pi$ in Fig.\ref{fig3}(a), $k$ changes from 0 to 1,1 to 2,2 to 3 at $\Delta_M= 500,100,1500 \times2\pi$ MHz, respectively. It can also be seen that the sensitivity remains constant over a range, because the best RF parameter $a/\omega$ takes the maximum value of the Bessel function $\mathbf{J}_k(a/\omega)$.

According to Fig.\ref{fig3}(a), the worst sensitivity in the extended frequency Range of 2GHz achieved by our scheme is only reduced by a factor of three times compared to resonance. Due to the abundance of Rydberg levels, we can find many pairs of Rydberg states that satisfy $a_{max}=1000\times2\pi$MHz and $\omega_{max}=500\times2\pi$MHz.  Given the best sensitivity achieved so far \cite{16-10.1063/5.0146768}, it is reasonable to assume that the optimal sensitivity based on resonance measurement is 20 $\mathrm{nV cm^{-1} Hz^{-\frac{1}{2}}}$. This implies that, theoretically, when the detuning is from 100MHz to 2GHz, the optimum sensitivity of our scheme is also 20 $\mathrm{nV cm^{-1} Hz^{-\frac{1}{2}}}$, and the worst sensitivity is astonishingly 54 $\mathrm{nV cm^{-1} Hz^{-\frac{1}{2}}}$, which is orders of magnitude better than other continuous frequency measurement schemes.

\section{conclusion \label{sec5}}
In summary, we have proposed a method for continuous MW field measurement utilizing quantum mixing technology in a Rydberg atom-based MW sensor, achieved by applying a driving control field. The control field, derived from a low-frequency weak radio-frequency (RF) field, induces an energy shift in the Rydberg state that oscillates with time due to the AC Stark effect. The distinct polarizabilities between two Rydberg states enable the control field and MW field to be non-commutative, fulfilling the requirements for quantum mixing. In our approach, the second-order effects arising from quantum mixing must be meticulously considered, as they can perturb the heterodyne signal and consequently compromise the accuracy of the electric field measurement.

When compared to resonance measurements, the sensitivity degradation for a 2.0 GHz far-detuned MW field is less than threefold, representing at least an order of magnitude improvement over alternative methods. We have demonstrated that by increasing the intensity and frequency of the RF field, the sensitivity of the far-detuned MW field can be substantially enhanced. Hence, future endeavors, particularly experimental implementations, should prioritize the selection of suitable Rydberg states that allow for the maximization of the RF field's frequency and intensity. Given the abundance of Rydberg levels, we anticipate achieving the worst sensitivity of 54 $\mathrm{nV cm^{-1} Hz^{-\frac{1}{2}}}$ within the extended frequency range of 100 MHz to 2 GHz. This work paves the way for high sensitivity and wide broadband MW field measurements using Rydberg atoms.

\begin{acknowledgments}
This work is supported by National Science Foundation of China under Grant No. 12274045, No. 12274046 and No.12347101.
\end{acknowledgments}

\appendix
\section{the theory of quantum frequency mixing \label{ap1}}
We consider a time-dependent Hamiltonian that can be described by two frequency modes $(\omega_a,\omega_b)$: 
\begin{eqnarray}
	H(t)=\sum_{m,n}H_{m,n}e^{im\omega_at}e^{im\omega_bt},\label{eq3}
\end{eqnarray}
where $\omega_a$ and $\omega_b$ are much larger than other energy parameters. According to the theory of the quantum frequency mixing, the time-dependent effective Hamiltonian with low-frequency can be obtained as \cite{25-PhysRevX.12.021061}:
\begin{eqnarray}
	H^{eff}(t)=\sum_{l,k}\left( H_{l,k}+H^{(2)}_{l,k}+\dots\right)e^{i(l\omega_a+k\omega_b)t},\label{eq4}
\end{eqnarray}
where summation indices $(l,k)$ satisfy the condition that $l\omega_a+k\omega_b$ is a low frequency. The second-order term $H^{(2)}_{l,k}$ is given by:
\begin{eqnarray}
	-\frac{1}{2}\sum_{(p,q)\neq (l,k)}\frac{\left[ H_{l-p,k-q},H_{p,q}\right] }{p\omega_a+q\omega_b}\label{eq5}
\end{eqnarray}
where the summation excludes the case $(p,q)=(l,k)$ to avoid divergency. Non-commutation in second-order terms enables quantum frequency mixing: the non-commutation between high-frequency terms is crucial to quantum mixing theory, implying that the applied controlled field must not commute with the signal field. This theory can be generalized to the case of multiple frequency modes.

\bibliographystyle{apsrev4-2}
\bibliography{ref}

\begin{thebibliography}{31}%
\makeatletter
\providecommand \@ifxundefined [1]{%
 \@ifx{#1\undefined}
}%
\providecommand \@ifnum [1]{%
 \ifnum #1\expandafter \@firstoftwo
 \else \expandafter \@secondoftwo
 \fi
}%
\providecommand \@ifx [1]{%
 \ifx #1\expandafter \@firstoftwo
 \else \expandafter \@secondoftwo
 \fi
}%
\providecommand \natexlab [1]{#1}%
\providecommand \enquote  [1]{``#1''}%
\providecommand \bibnamefont  [1]{#1}%
\providecommand \bibfnamefont [1]{#1}%
\providecommand \citenamefont [1]{#1}%
\providecommand \href@noop [0]{\@secondoftwo}%
\providecommand \href [0]{\begingroup \@sanitize@url \@href}%
\providecommand \@href[1]{\@@startlink{#1}\@@href}%
\providecommand \@@href[1]{\endgroup#1\@@endlink}%
\providecommand \@sanitize@url [0]{\catcode `\\12\catcode `\$12\catcode
  `\&12\catcode `\#12\catcode `\^12\catcode `\_12\catcode `\%12\relax}%
\providecommand \@@startlink[1]{}%
\providecommand \@@endlink[0]{}%
\providecommand \url  [0]{\begingroup\@sanitize@url \@url }%
\providecommand \@url [1]{\endgroup\@href {#1}{\urlprefix }}%
\providecommand \urlprefix  [0]{URL }%
\providecommand \Eprint [0]{\href }%
\providecommand \doibase [0]{https://doi.org/}%
\providecommand \selectlanguage [0]{\@gobble}%
\providecommand \bibinfo  [0]{\@secondoftwo}%
\providecommand \bibfield  [0]{\@secondoftwo}%
\providecommand \translation [1]{[#1]}%
\providecommand \BibitemOpen [0]{}%
\providecommand \bibitemStop [0]{}%
\providecommand \bibitemNoStop [0]{.\EOS\space}%
\providecommand \EOS [0]{\spacefactor3000\relax}%
\providecommand \BibitemShut  [1]{\csname bibitem#1\endcsname}%
\let\auto@bib@innerbib\@empty
%</preamble>
\bibitem [{\citenamefont {Fan}\ \emph {et~al.}(2015)\citenamefont {Fan},
  \citenamefont {Kumar}, \citenamefont {Sedlacek}, \citenamefont {Kübler},
  \citenamefont {Karimkashi},\ and\ \citenamefont {Shaffer}}]{1-Fan_2015}%
  \BibitemOpen
  \bibfield  {author} {\bibinfo {author} {\bibfnamefont {H.}~\bibnamefont
  {Fan}}, \bibinfo {author} {\bibfnamefont {S.}~\bibnamefont {Kumar}}, \bibinfo
  {author} {\bibfnamefont {J.}~\bibnamefont {Sedlacek}}, \bibinfo {author}
  {\bibfnamefont {H.}~\bibnamefont {Kübler}}, \bibinfo {author} {\bibfnamefont
  {S.}~\bibnamefont {Karimkashi}},\ and\ \bibinfo {author} {\bibfnamefont
  {J.~P.}\ \bibnamefont {Shaffer}},\ }\href
  {https://doi.org/10.1088/0953-4075/48/20/202001} {\bibfield  {journal}
  {\bibinfo  {journal} {Journal of Physics B: Atomic, Molecular and Optical
  Physics}\ }\textbf {\bibinfo {volume} {48}},\ \bibinfo {pages} {202001}
  (\bibinfo {year} {2015})}\BibitemShut {NoStop}%
\bibitem [{\citenamefont {Yuan}\ \emph {et~al.}(2023)\citenamefont {Yuan},
  \citenamefont {Yang}, \citenamefont {Jing}, \citenamefont {Zhang},
  \citenamefont {Jiao}, \citenamefont {Li}, \citenamefont {Zhang},
  \citenamefont {Xiao},\ and\ \citenamefont {Jia}}]{2-Yuan_2023}%
  \BibitemOpen
  \bibfield  {author} {\bibinfo {author} {\bibfnamefont {J.}~\bibnamefont
  {Yuan}}, \bibinfo {author} {\bibfnamefont {W.}~\bibnamefont {Yang}}, \bibinfo
  {author} {\bibfnamefont {M.}~\bibnamefont {Jing}}, \bibinfo {author}
  {\bibfnamefont {H.}~\bibnamefont {Zhang}}, \bibinfo {author} {\bibfnamefont
  {Y.}~\bibnamefont {Jiao}}, \bibinfo {author} {\bibfnamefont {W.}~\bibnamefont
  {Li}}, \bibinfo {author} {\bibfnamefont {L.}~\bibnamefont {Zhang}}, \bibinfo
  {author} {\bibfnamefont {L.}~\bibnamefont {Xiao}},\ and\ \bibinfo {author}
  {\bibfnamefont {S.}~\bibnamefont {Jia}},\ }\href
  {https://doi.org/10.1088/1361-6633/acf22f} {\bibfield  {journal} {\bibinfo
  {journal} {Reports on Progress in Physics}\ }\textbf {\bibinfo {volume}
  {86}},\ \bibinfo {pages} {106001} (\bibinfo {year} {2023})}\BibitemShut
  {NoStop}%
\bibitem [{\citenamefont {Zhang}\ \emph {et~al.}(2024)\citenamefont {Zhang},
  \citenamefont {Ma}, \citenamefont {Liao}, \citenamefont {Yang}, \citenamefont
  {Liu}, \citenamefont {Ding}, \citenamefont {Yan}, \citenamefont {Li},\ and\
  \citenamefont {Zhang}}]{3-ZHANG20241515}%
  \BibitemOpen
  \bibfield  {author} {\bibinfo {author} {\bibfnamefont {H.}~\bibnamefont
  {Zhang}}, \bibinfo {author} {\bibfnamefont {Y.}~\bibnamefont {Ma}}, \bibinfo
  {author} {\bibfnamefont {K.}~\bibnamefont {Liao}}, \bibinfo {author}
  {\bibfnamefont {W.}~\bibnamefont {Yang}}, \bibinfo {author} {\bibfnamefont
  {Z.}~\bibnamefont {Liu}}, \bibinfo {author} {\bibfnamefont {D.}~\bibnamefont
  {Ding}}, \bibinfo {author} {\bibfnamefont {H.}~\bibnamefont {Yan}}, \bibinfo
  {author} {\bibfnamefont {W.}~\bibnamefont {Li}},\ and\ \bibinfo {author}
  {\bibfnamefont {L.}~\bibnamefont {Zhang}},\ }\href
  {https://doi.org/https://doi.org/10.1016/j.scib.2024.03.032} {\bibfield
  {journal} {\bibinfo  {journal} {Science Bulletin}\ }\textbf {\bibinfo
  {volume} {69}},\ \bibinfo {pages} {1515} (\bibinfo {year}
  {2024})}\BibitemShut {NoStop}%
\bibitem [{\citenamefont {Fleischhauer}\ \emph {et~al.}(2005)\citenamefont
  {Fleischhauer}, \citenamefont {Imamoglu},\ and\ \citenamefont
  {Marangos}}]{4-RevModPhys.77.633}%
  \BibitemOpen
  \bibfield  {author} {\bibinfo {author} {\bibfnamefont {M.}~\bibnamefont
  {Fleischhauer}}, \bibinfo {author} {\bibfnamefont {A.}~\bibnamefont
  {Imamoglu}},\ and\ \bibinfo {author} {\bibfnamefont {J.~P.}\ \bibnamefont
  {Marangos}},\ }\href {https://doi.org/10.1103/RevModPhys.77.633} {\bibfield
  {journal} {\bibinfo  {journal} {Rev. Mod. Phys.}\ }\textbf {\bibinfo {volume}
  {77}},\ \bibinfo {pages} {633} (\bibinfo {year} {2005})}\BibitemShut
  {NoStop}%
\bibitem [{\citenamefont {Abi-Salloum}(2010)}]{5-PhysRevA.81.053836}%
  \BibitemOpen
  \bibfield  {author} {\bibinfo {author} {\bibfnamefont {T.~Y.}\ \bibnamefont
  {Abi-Salloum}},\ }\href {https://doi.org/10.1103/PhysRevA.81.053836}
  {\bibfield  {journal} {\bibinfo  {journal} {Phys. Rev. A}\ }\textbf {\bibinfo
  {volume} {81}},\ \bibinfo {pages} {053836} (\bibinfo {year}
  {2010})}\BibitemShut {NoStop}%
\bibitem [{\citenamefont {Deb}\ and\ \citenamefont
  {Kjærgaard}(2018)}]{6-10.1063/1.5031033}%
  \BibitemOpen
  \bibfield  {author} {\bibinfo {author} {\bibfnamefont {A.~B.}\ \bibnamefont
  {Deb}}\ and\ \bibinfo {author} {\bibfnamefont {N.}~\bibnamefont
  {Kjærgaard}},\ }\href {https://doi.org/10.1063/1.5031033} {\bibfield
  {journal} {\bibinfo  {journal} {Applied Physics Letters}\ }\textbf {\bibinfo
  {volume} {112}},\ \bibinfo {pages} {211106} (\bibinfo {year}
  {2018})}\BibitemShut {NoStop}%
\bibitem [{\citenamefont {Song}\ \emph {et~al.}(2019)\citenamefont {Song},
  \citenamefont {Liu}, \citenamefont {Liu}, \citenamefont {Zhang},
  \citenamefont {Zou}, \citenamefont {Zhang},\ and\ \citenamefont
  {Qu}}]{7-Song:19}%
  \BibitemOpen
  \bibfield  {author} {\bibinfo {author} {\bibfnamefont {Z.}~\bibnamefont
  {Song}}, \bibinfo {author} {\bibfnamefont {H.}~\bibnamefont {Liu}}, \bibinfo
  {author} {\bibfnamefont {X.}~\bibnamefont {Liu}}, \bibinfo {author}
  {\bibfnamefont {W.}~\bibnamefont {Zhang}}, \bibinfo {author} {\bibfnamefont
  {H.}~\bibnamefont {Zou}}, \bibinfo {author} {\bibfnamefont {J.}~\bibnamefont
  {Zhang}},\ and\ \bibinfo {author} {\bibfnamefont {J.}~\bibnamefont {Qu}},\
  }\href {https://doi.org/10.1364/OE.27.008848} {\bibfield  {journal} {\bibinfo
   {journal} {Opt. Express}\ }\textbf {\bibinfo {volume} {27}},\ \bibinfo
  {pages} {8848} (\bibinfo {year} {2019})}\BibitemShut {NoStop}%
\bibitem [{\citenamefont {Anderson}\ \emph {et~al.}(2021)\citenamefont
  {Anderson}, \citenamefont {Sapiro},\ and\ \citenamefont
  {Raithel}}]{8-9069423}%
  \BibitemOpen
  \bibfield  {author} {\bibinfo {author} {\bibfnamefont {D.~A.}\ \bibnamefont
  {Anderson}}, \bibinfo {author} {\bibfnamefont {R.~E.}\ \bibnamefont
  {Sapiro}},\ and\ \bibinfo {author} {\bibfnamefont {G.}~\bibnamefont
  {Raithel}},\ }\href {https://doi.org/10.1109/TAP.2020.2987112} {\bibfield
  {journal} {\bibinfo  {journal} {IEEE Transactions on Antennas and
  Propagation}\ }\textbf {\bibinfo {volume} {69}},\ \bibinfo {pages} {2455}
  (\bibinfo {year} {2021})}\BibitemShut {NoStop}%
\bibitem [{\citenamefont {Sedlacek}\ \emph {et~al.}(2013)\citenamefont
  {Sedlacek}, \citenamefont {Schwettmann}, \citenamefont {K\"ubler},\ and\
  \citenamefont {Shaffer}}]{9-PhysRevLett.111.063001}%
  \BibitemOpen
  \bibfield  {author} {\bibinfo {author} {\bibfnamefont {J.~A.}\ \bibnamefont
  {Sedlacek}}, \bibinfo {author} {\bibfnamefont {A.}~\bibnamefont
  {Schwettmann}}, \bibinfo {author} {\bibfnamefont {H.}~\bibnamefont
  {K\"ubler}},\ and\ \bibinfo {author} {\bibfnamefont {J.~P.}\ \bibnamefont
  {Shaffer}},\ }\href {https://doi.org/10.1103/PhysRevLett.111.063001}
  {\bibfield  {journal} {\bibinfo  {journal} {Phys. Rev. Lett.}\ }\textbf
  {\bibinfo {volume} {111}},\ \bibinfo {pages} {063001} (\bibinfo {year}
  {2013})}\BibitemShut {NoStop}%
\bibitem [{\citenamefont {Fan}\ \emph {et~al.}(2014)\citenamefont {Fan},
  \citenamefont {Kumar}, \citenamefont {Daschner}, \citenamefont {K\"{u}bler},\
  and\ \citenamefont {Shaffer}}]{10-Fan:14}%
  \BibitemOpen
  \bibfield  {author} {\bibinfo {author} {\bibfnamefont {H.~Q.}\ \bibnamefont
  {Fan}}, \bibinfo {author} {\bibfnamefont {S.}~\bibnamefont {Kumar}}, \bibinfo
  {author} {\bibfnamefont {R.}~\bibnamefont {Daschner}}, \bibinfo {author}
  {\bibfnamefont {H.}~\bibnamefont {K\"{u}bler}},\ and\ \bibinfo {author}
  {\bibfnamefont {J.~P.}\ \bibnamefont {Shaffer}},\ }\href
  {https://doi.org/10.1364/OL.39.003030} {\bibfield  {journal} {\bibinfo
  {journal} {Opt. Lett.}\ }\textbf {\bibinfo {volume} {39}},\ \bibinfo {pages}
  {3030} (\bibinfo {year} {2014})}\BibitemShut {NoStop}%
\bibitem [{\citenamefont {Sedlacek}\ \emph {et~al.}(2012)\citenamefont
  {Sedlacek}, \citenamefont {Schwettmann}, \citenamefont {Kübler},
  \citenamefont {Löw}, \citenamefont {Pfau},\ and\ \citenamefont
  {Shaffer}}]{11-Sedlacek}%
  \BibitemOpen
  \bibfield  {author} {\bibinfo {author} {\bibfnamefont {J.~A.}\ \bibnamefont
  {Sedlacek}}, \bibinfo {author} {\bibfnamefont {A.}~\bibnamefont
  {Schwettmann}}, \bibinfo {author} {\bibfnamefont {H.}~\bibnamefont
  {Kübler}}, \bibinfo {author} {\bibfnamefont {R.}~\bibnamefont {Löw}},
  \bibinfo {author} {\bibfnamefont {T.}~\bibnamefont {Pfau}},\ and\ \bibinfo
  {author} {\bibfnamefont {J.~P.}\ \bibnamefont {Shaffer}},\ }\href
  {https://doi.org/10.1038/nphys2423} {\bibfield  {journal} {\bibinfo
  {journal} {Nature Physics}\ }\textbf {\bibinfo {volume} {8}},\ \bibinfo
  {pages} {819–824} (\bibinfo {year} {2012})}\BibitemShut {NoStop}%
\bibitem [{\citenamefont {Jing}\ \emph {et~al.}(2020)\citenamefont {Jing},
  \citenamefont {Hu}, \citenamefont {Ma}, \citenamefont {Zhang}, \citenamefont
  {Zhang}, \citenamefont {Xiao},\ and\ \citenamefont {Jia}}]{12-Jing}%
  \BibitemOpen
  \bibfield  {author} {\bibinfo {author} {\bibfnamefont {M.}~\bibnamefont
  {Jing}}, \bibinfo {author} {\bibfnamefont {Y.}~\bibnamefont {Hu}}, \bibinfo
  {author} {\bibfnamefont {J.}~\bibnamefont {Ma}}, \bibinfo {author}
  {\bibfnamefont {H.}~\bibnamefont {Zhang}}, \bibinfo {author} {\bibfnamefont
  {L.}~\bibnamefont {Zhang}}, \bibinfo {author} {\bibfnamefont
  {L.}~\bibnamefont {Xiao}},\ and\ \bibinfo {author} {\bibfnamefont
  {S.}~\bibnamefont {Jia}},\ }\href {https://doi.org/10.1038/s41567-020-0918-5}
  {\bibfield  {journal} {\bibinfo  {journal} {Nature Physics}\ }\textbf
  {\bibinfo {volume} {16}},\ \bibinfo {pages} {911–915} (\bibinfo {year}
  {2020})}\BibitemShut {NoStop}%
\bibitem [{\citenamefont {Simons}\ \emph {et~al.}(2016)\citenamefont {Simons},
  \citenamefont {Gordon}, \citenamefont {Holloway}, \citenamefont {Anderson},
  \citenamefont {Miller},\ and\ \citenamefont
  {Raithel}}]{13-10.1063/1.4947231}%
  \BibitemOpen
  \bibfield  {author} {\bibinfo {author} {\bibfnamefont {M.~T.}\ \bibnamefont
  {Simons}}, \bibinfo {author} {\bibfnamefont {J.~A.}\ \bibnamefont {Gordon}},
  \bibinfo {author} {\bibfnamefont {C.~L.}\ \bibnamefont {Holloway}}, \bibinfo
  {author} {\bibfnamefont {D.~A.}\ \bibnamefont {Anderson}}, \bibinfo {author}
  {\bibfnamefont {S.~A.}\ \bibnamefont {Miller}},\ and\ \bibinfo {author}
  {\bibfnamefont {G.}~\bibnamefont {Raithel}},\ }\href
  {https://doi.org/10.1063/1.4947231} {\bibfield  {journal} {\bibinfo
  {journal} {Applied Physics Letters}\ }\textbf {\bibinfo {volume} {108}},\
  \bibinfo {pages} {174101} (\bibinfo {year} {2016})}\BibitemShut {NoStop}%
\bibitem [{\citenamefont {Kumar}\ \emph {et~al.}(2017)\citenamefont {Kumar},
  \citenamefont {Fan}, \citenamefont {Kübler}, \citenamefont {Sheng},\ and\
  \citenamefont {Shaffer}}]{14-Kumar}%
  \BibitemOpen
  \bibfield  {author} {\bibinfo {author} {\bibfnamefont {S.}~\bibnamefont
  {Kumar}}, \bibinfo {author} {\bibfnamefont {H.}~\bibnamefont {Fan}}, \bibinfo
  {author} {\bibfnamefont {H.}~\bibnamefont {Kübler}}, \bibinfo {author}
  {\bibfnamefont {J.}~\bibnamefont {Sheng}},\ and\ \bibinfo {author}
  {\bibfnamefont {J.~P.}\ \bibnamefont {Shaffer}},\ }\href
  {https://doi.org/10.1038/srep42981} {\bibfield  {journal} {\bibinfo
  {journal} {Scientific Reports}\ }\textbf {\bibinfo {volume} {7}},\ \bibinfo
  {pages} {42981} (\bibinfo {year} {2017})}\BibitemShut {NoStop}%
\bibitem [{\citenamefont {Prajapati}\ \emph {et~al.}(2021)\citenamefont
  {Prajapati}, \citenamefont {Robinson}, \citenamefont {Berweger},
  \citenamefont {Simons}, \citenamefont {Artusio-Glimpse},\ and\ \citenamefont
  {Holloway}}]{15-10.1063/5.0069195}%
  \BibitemOpen
  \bibfield  {author} {\bibinfo {author} {\bibfnamefont {N.}~\bibnamefont
  {Prajapati}}, \bibinfo {author} {\bibfnamefont {A.~K.}\ \bibnamefont
  {Robinson}}, \bibinfo {author} {\bibfnamefont {S.}~\bibnamefont {Berweger}},
  \bibinfo {author} {\bibfnamefont {M.~T.}\ \bibnamefont {Simons}}, \bibinfo
  {author} {\bibfnamefont {A.~B.}\ \bibnamefont {Artusio-Glimpse}},\ and\
  \bibinfo {author} {\bibfnamefont {C.~L.}\ \bibnamefont {Holloway}},\ }\href
  {https://doi.org/10.1063/5.0069195} {\bibfield  {journal} {\bibinfo
  {journal} {Applied Physics Letters}\ }\textbf {\bibinfo {volume} {119}},\
  \bibinfo {pages} {214001} (\bibinfo {year} {2021})}\BibitemShut {NoStop}%
\bibitem [{\citenamefont {Cai}\ \emph {et~al.}(2023)\citenamefont {Cai},
  \citenamefont {You}, \citenamefont {Zhang}, \citenamefont {Xu},\ and\
  \citenamefont {Liu}}]{16-10.1063/5.0146768}%
  \BibitemOpen
  \bibfield  {author} {\bibinfo {author} {\bibfnamefont {M.}~\bibnamefont
  {Cai}}, \bibinfo {author} {\bibfnamefont {S.}~\bibnamefont {You}}, \bibinfo
  {author} {\bibfnamefont {S.}~\bibnamefont {Zhang}}, \bibinfo {author}
  {\bibfnamefont {Z.}~\bibnamefont {Xu}},\ and\ \bibinfo {author}
  {\bibfnamefont {H.}~\bibnamefont {Liu}},\ }\href
  {https://doi.org/10.1063/5.0146768} {\bibfield  {journal} {\bibinfo
  {journal} {Applied Physics Letters}\ }\textbf {\bibinfo {volume} {122}},\
  \bibinfo {pages} {161103} (\bibinfo {year} {2023})}\BibitemShut {NoStop}%
\bibitem [{\citenamefont {Meyer}\ \emph {et~al.}(2021)\citenamefont {Meyer},
  \citenamefont {Kunz},\ and\ \citenamefont
  {Cox}}]{17-PhysRevApplied.15.014053}%
  \BibitemOpen
  \bibfield  {author} {\bibinfo {author} {\bibfnamefont {D.~H.}\ \bibnamefont
  {Meyer}}, \bibinfo {author} {\bibfnamefont {P.~D.}\ \bibnamefont {Kunz}},\
  and\ \bibinfo {author} {\bibfnamefont {K.~C.}\ \bibnamefont {Cox}},\ }\href
  {https://doi.org/10.1103/PhysRevApplied.15.014053} {\bibfield  {journal}
  {\bibinfo  {journal} {Phys. Rev. Appl.}\ }\textbf {\bibinfo {volume} {15}},\
  \bibinfo {pages} {014053} (\bibinfo {year} {2021})}\BibitemShut {NoStop}%
\bibitem [{\citenamefont {Hu}\ \emph {et~al.}(2022)\citenamefont {Hu},
  \citenamefont {Li}, \citenamefont {Song}, \citenamefont {Bai}, \citenamefont
  {Jiao}, \citenamefont {Zhao},\ and\ \citenamefont
  {Jia}}]{18-10.1063/5.0086357}%
  \BibitemOpen
  \bibfield  {author} {\bibinfo {author} {\bibfnamefont {J.}~\bibnamefont
  {Hu}}, \bibinfo {author} {\bibfnamefont {H.}~\bibnamefont {Li}}, \bibinfo
  {author} {\bibfnamefont {R.}~\bibnamefont {Song}}, \bibinfo {author}
  {\bibfnamefont {J.}~\bibnamefont {Bai}}, \bibinfo {author} {\bibfnamefont
  {Y.}~\bibnamefont {Jiao}}, \bibinfo {author} {\bibfnamefont {J.}~\bibnamefont
  {Zhao}},\ and\ \bibinfo {author} {\bibfnamefont {S.}~\bibnamefont {Jia}},\
  }\href {https://doi.org/10.1063/5.0086357} {\bibfield  {journal} {\bibinfo
  {journal} {Applied Physics Letters}\ }\textbf {\bibinfo {volume} {121}},\
  \bibinfo {pages} {014002} (\bibinfo {year} {2022})}\BibitemShut {NoStop}%
\bibitem [{\citenamefont {Simons}\ \emph {et~al.}(2021)\citenamefont {Simons},
  \citenamefont {Artusio-Glimpse}, \citenamefont {Holloway}, \citenamefont
  {Imhof}, \citenamefont {Jefferts}, \citenamefont {Wyllie}, \citenamefont
  {Sawyer},\ and\ \citenamefont {Walker}}]{19-PhysRevA.104.032824}%
  \BibitemOpen
  \bibfield  {author} {\bibinfo {author} {\bibfnamefont {M.~T.}\ \bibnamefont
  {Simons}}, \bibinfo {author} {\bibfnamefont {A.~B.}\ \bibnamefont
  {Artusio-Glimpse}}, \bibinfo {author} {\bibfnamefont {C.~L.}\ \bibnamefont
  {Holloway}}, \bibinfo {author} {\bibfnamefont {E.}~\bibnamefont {Imhof}},
  \bibinfo {author} {\bibfnamefont {S.~R.}\ \bibnamefont {Jefferts}}, \bibinfo
  {author} {\bibfnamefont {R.}~\bibnamefont {Wyllie}}, \bibinfo {author}
  {\bibfnamefont {B.~C.}\ \bibnamefont {Sawyer}},\ and\ \bibinfo {author}
  {\bibfnamefont {T.~G.}\ \bibnamefont {Walker}},\ }\href
  {https://doi.org/10.1103/PhysRevA.104.032824} {\bibfield  {journal} {\bibinfo
   {journal} {Phys. Rev. A}\ }\textbf {\bibinfo {volume} {104}},\ \bibinfo
  {pages} {032824} (\bibinfo {year} {2021})}\BibitemShut {NoStop}%
\bibitem [{\citenamefont {Berweger}\ \emph {et~al.}(2023)\citenamefont
  {Berweger}, \citenamefont {Prajapati}, \citenamefont {Artusio-Glimpse},
  \citenamefont {Rotunno}, \citenamefont {Brown}, \citenamefont {Holloway},
  \citenamefont {Simons}, \citenamefont {Imhof}, \citenamefont {Jefferts},
  \citenamefont {Kayim}, \citenamefont {Viray}, \citenamefont {Wyllie},
  \citenamefont {Sawyer},\ and\ \citenamefont
  {Walker}}]{20-PhysRevApplied.19.044049}%
  \BibitemOpen
  \bibfield  {author} {\bibinfo {author} {\bibfnamefont {S.}~\bibnamefont
  {Berweger}}, \bibinfo {author} {\bibfnamefont {N.}~\bibnamefont {Prajapati}},
  \bibinfo {author} {\bibfnamefont {A.~B.}\ \bibnamefont {Artusio-Glimpse}},
  \bibinfo {author} {\bibfnamefont {A.~P.}\ \bibnamefont {Rotunno}}, \bibinfo
  {author} {\bibfnamefont {R.}~\bibnamefont {Brown}}, \bibinfo {author}
  {\bibfnamefont {C.~L.}\ \bibnamefont {Holloway}}, \bibinfo {author}
  {\bibfnamefont {M.~T.}\ \bibnamefont {Simons}}, \bibinfo {author}
  {\bibfnamefont {E.}~\bibnamefont {Imhof}}, \bibinfo {author} {\bibfnamefont
  {S.~R.}\ \bibnamefont {Jefferts}}, \bibinfo {author} {\bibfnamefont {B.~N.}\
  \bibnamefont {Kayim}}, \bibinfo {author} {\bibfnamefont {M.~A.}\ \bibnamefont
  {Viray}}, \bibinfo {author} {\bibfnamefont {R.}~\bibnamefont {Wyllie}},
  \bibinfo {author} {\bibfnamefont {B.~C.}\ \bibnamefont {Sawyer}},\ and\
  \bibinfo {author} {\bibfnamefont {T.~G.}\ \bibnamefont {Walker}},\ }\href
  {https://doi.org/10.1103/PhysRevApplied.19.044049} {\bibfield  {journal}
  {\bibinfo  {journal} {Phys. Rev. Appl.}\ }\textbf {\bibinfo {volume} {19}},\
  \bibinfo {pages} {044049} (\bibinfo {year} {2023})}\BibitemShut {NoStop}%
\bibitem [{\citenamefont {Anderson}\ and\ \citenamefont
  {Raithel}(2017)}]{21-10.1063/1.4996234}%
  \BibitemOpen
  \bibfield  {author} {\bibinfo {author} {\bibfnamefont {D.~A.}\ \bibnamefont
  {Anderson}}\ and\ \bibinfo {author} {\bibfnamefont {G.}~\bibnamefont
  {Raithel}},\ }\href {https://doi.org/10.1063/1.4996234} {\bibfield  {journal}
  {\bibinfo  {journal} {Applied Physics Letters}\ }\textbf {\bibinfo {volume}
  {111}},\ \bibinfo {pages} {053504} (\bibinfo {year} {2017})}\BibitemShut
  {NoStop}%
\bibitem [{\citenamefont {Liu}\ \emph {et~al.}(2022)\citenamefont {Liu},
  \citenamefont {Liao}, \citenamefont {Zhang}, \citenamefont {Tu},
  \citenamefont {Bian}, \citenamefont {Li}, \citenamefont {Zheng},
  \citenamefont {Li}, \citenamefont {Huang}, \citenamefont {Yan},\ and\
  \citenamefont {Zhu}}]{22-PhysRevApplied.18.054003}%
  \BibitemOpen
  \bibfield  {author} {\bibinfo {author} {\bibfnamefont {X.-H.}\ \bibnamefont
  {Liu}}, \bibinfo {author} {\bibfnamefont {K.-Y.}\ \bibnamefont {Liao}},
  \bibinfo {author} {\bibfnamefont {Z.-X.}\ \bibnamefont {Zhang}}, \bibinfo
  {author} {\bibfnamefont {H.-T.}\ \bibnamefont {Tu}}, \bibinfo {author}
  {\bibfnamefont {W.}~\bibnamefont {Bian}}, \bibinfo {author} {\bibfnamefont
  {Z.-Q.}\ \bibnamefont {Li}}, \bibinfo {author} {\bibfnamefont {S.-Y.}\
  \bibnamefont {Zheng}}, \bibinfo {author} {\bibfnamefont {H.-H.}\ \bibnamefont
  {Li}}, \bibinfo {author} {\bibfnamefont {W.}~\bibnamefont {Huang}}, \bibinfo
  {author} {\bibfnamefont {H.}~\bibnamefont {Yan}},\ and\ \bibinfo {author}
  {\bibfnamefont {S.-L.}\ \bibnamefont {Zhu}},\ }\href
  {https://doi.org/10.1103/PhysRevApplied.18.054003} {\bibfield  {journal}
  {\bibinfo  {journal} {Phys. Rev. Appl.}\ }\textbf {\bibinfo {volume} {18}},\
  \bibinfo {pages} {054003} (\bibinfo {year} {2022})}\BibitemShut {NoStop}%
\bibitem [{\citenamefont {Jia}\ \emph {et~al.}(2021)\citenamefont {Jia},
  \citenamefont {Liu}, \citenamefont {Mei}, \citenamefont {Yu}, \citenamefont
  {Zhang}, \citenamefont {Lin}, \citenamefont {Dong}, \citenamefont {Zhang},
  \citenamefont {Xie},\ and\ \citenamefont {Zhong}}]{23-PhysRevA.103.063113}%
  \BibitemOpen
  \bibfield  {author} {\bibinfo {author} {\bibfnamefont {F.-D.}\ \bibnamefont
  {Jia}}, \bibinfo {author} {\bibfnamefont {X.-B.}\ \bibnamefont {Liu}},
  \bibinfo {author} {\bibfnamefont {J.}~\bibnamefont {Mei}}, \bibinfo {author}
  {\bibfnamefont {Y.-H.}\ \bibnamefont {Yu}}, \bibinfo {author} {\bibfnamefont
  {H.-Y.}\ \bibnamefont {Zhang}}, \bibinfo {author} {\bibfnamefont {Z.-Q.}\
  \bibnamefont {Lin}}, \bibinfo {author} {\bibfnamefont {H.-Y.}\ \bibnamefont
  {Dong}}, \bibinfo {author} {\bibfnamefont {J.}~\bibnamefont {Zhang}},
  \bibinfo {author} {\bibfnamefont {F.}~\bibnamefont {Xie}},\ and\ \bibinfo
  {author} {\bibfnamefont {Z.-P.}\ \bibnamefont {Zhong}},\ }\href
  {https://doi.org/10.1103/PhysRevA.103.063113} {\bibfield  {journal} {\bibinfo
   {journal} {Phys. Rev. A}\ }\textbf {\bibinfo {volume} {103}},\ \bibinfo
  {pages} {063113} (\bibinfo {year} {2021})}\BibitemShut {NoStop}%
\bibitem [{\citenamefont {Cui}\ \emph {et~al.}(2023)\citenamefont {Cui},
  \citenamefont {Jia}, \citenamefont {Hao}, \citenamefont {Wang}, \citenamefont
  {Zhou}, \citenamefont {Liu}, \citenamefont {Yu}, \citenamefont {Mei},
  \citenamefont {Bai}, \citenamefont {Bao}, \citenamefont {Hu}, \citenamefont
  {Wang}, \citenamefont {Liu}, \citenamefont {Zhang}, \citenamefont {Xie},\
  and\ \citenamefont {Zhong}}]{24-PhysRevA.107.043102}%
  \BibitemOpen
  \bibfield  {author} {\bibinfo {author} {\bibfnamefont {Y.}~\bibnamefont
  {Cui}}, \bibinfo {author} {\bibfnamefont {F.-D.}\ \bibnamefont {Jia}},
  \bibinfo {author} {\bibfnamefont {J.-H.}\ \bibnamefont {Hao}}, \bibinfo
  {author} {\bibfnamefont {Y.-H.}\ \bibnamefont {Wang}}, \bibinfo {author}
  {\bibfnamefont {F.}~\bibnamefont {Zhou}}, \bibinfo {author} {\bibfnamefont
  {X.-B.}\ \bibnamefont {Liu}}, \bibinfo {author} {\bibfnamefont {Y.-H.}\
  \bibnamefont {Yu}}, \bibinfo {author} {\bibfnamefont {J.}~\bibnamefont
  {Mei}}, \bibinfo {author} {\bibfnamefont {J.-H.}\ \bibnamefont {Bai}},
  \bibinfo {author} {\bibfnamefont {Y.-Y.}\ \bibnamefont {Bao}}, \bibinfo
  {author} {\bibfnamefont {D.}~\bibnamefont {Hu}}, \bibinfo {author}
  {\bibfnamefont {Y.}~\bibnamefont {Wang}}, \bibinfo {author} {\bibfnamefont
  {Y.}~\bibnamefont {Liu}}, \bibinfo {author} {\bibfnamefont {J.}~\bibnamefont
  {Zhang}}, \bibinfo {author} {\bibfnamefont {F.}~\bibnamefont {Xie}},\ and\
  \bibinfo {author} {\bibfnamefont {Z.-P.}\ \bibnamefont {Zhong}},\ }\href
  {https://doi.org/10.1103/PhysRevA.107.043102} {\bibfield  {journal} {\bibinfo
   {journal} {Phys. Rev. A}\ }\textbf {\bibinfo {volume} {107}},\ \bibinfo
  {pages} {043102} (\bibinfo {year} {2023})}\BibitemShut {NoStop}%
\bibitem [{\citenamefont {Wang}\ \emph {et~al.}(2022)\citenamefont {Wang},
  \citenamefont {Liu}, \citenamefont {Schloss}, \citenamefont {Alsid},
  \citenamefont {Braje},\ and\ \citenamefont
  {Cappellaro}}]{25-PhysRevX.12.021061}%
  \BibitemOpen
  \bibfield  {author} {\bibinfo {author} {\bibfnamefont {G.}~\bibnamefont
  {Wang}}, \bibinfo {author} {\bibfnamefont {Y.-X.}\ \bibnamefont {Liu}},
  \bibinfo {author} {\bibfnamefont {J.~M.}\ \bibnamefont {Schloss}}, \bibinfo
  {author} {\bibfnamefont {S.~T.}\ \bibnamefont {Alsid}}, \bibinfo {author}
  {\bibfnamefont {D.~A.}\ \bibnamefont {Braje}},\ and\ \bibinfo {author}
  {\bibfnamefont {P.}~\bibnamefont {Cappellaro}},\ }\href
  {https://doi.org/10.1103/PhysRevX.12.021061} {\bibfield  {journal} {\bibinfo
  {journal} {Phys. Rev. X}\ }\textbf {\bibinfo {volume} {12}},\ \bibinfo
  {pages} {021061} (\bibinfo {year} {2022})}\BibitemShut {NoStop}%
\bibitem [{\citenamefont {Bason}\ \emph {et~al.}(2010)\citenamefont {Bason},
  \citenamefont {Tanasittikosol}, \citenamefont {Sargsyan}, \citenamefont
  {Mohapatra}, \citenamefont {Sarkisyan}, \citenamefont {Potvliege},\ and\
  \citenamefont {Adams}}]{26-Bason_2010}%
  \BibitemOpen
  \bibfield  {author} {\bibinfo {author} {\bibfnamefont {M.~G.}\ \bibnamefont
  {Bason}}, \bibinfo {author} {\bibfnamefont {M.}~\bibnamefont
  {Tanasittikosol}}, \bibinfo {author} {\bibfnamefont {A.}~\bibnamefont
  {Sargsyan}}, \bibinfo {author} {\bibfnamefont {A.~K.}\ \bibnamefont
  {Mohapatra}}, \bibinfo {author} {\bibfnamefont {D.}~\bibnamefont
  {Sarkisyan}}, \bibinfo {author} {\bibfnamefont {R.~M.}\ \bibnamefont
  {Potvliege}},\ and\ \bibinfo {author} {\bibfnamefont {C.~S.}\ \bibnamefont
  {Adams}},\ }\href {https://doi.org/10.1088/1367-2630/12/6/065015} {\bibfield
  {journal} {\bibinfo  {journal} {New Journal of Physics}\ }\textbf {\bibinfo
  {volume} {12}},\ \bibinfo {pages} {065015} (\bibinfo {year}
  {2010})}\BibitemShut {NoStop}%
\bibitem [{\citenamefont {Liu}\ \emph {et~al.}(2023)\citenamefont {Liu},
  \citenamefont {Zhang},\ and\ \citenamefont {Wang}}]{27-Liu_2023}%
  \BibitemOpen
  \bibfield  {author} {\bibinfo {author} {\bibfnamefont {W.}~\bibnamefont
  {Liu}}, \bibinfo {author} {\bibfnamefont {L.}~\bibnamefont {Zhang}},\ and\
  \bibinfo {author} {\bibfnamefont {T.}~\bibnamefont {Wang}},\ }\href
  {https://doi.org/10.1088/1674-1056/aca6db} {\bibfield  {journal} {\bibinfo
  {journal} {Chinese Physics B}\ }\textbf {\bibinfo {volume} {32}},\ \bibinfo
  {pages} {053203} (\bibinfo {year} {2023})}\BibitemShut {NoStop}%
\bibitem [{\citenamefont {Miller}\ \emph {et~al.}(2016)\citenamefont {Miller},
  \citenamefont {Anderson},\ and\ \citenamefont {Raithel}}]{28-Miller_2016}%
  \BibitemOpen
  \bibfield  {author} {\bibinfo {author} {\bibfnamefont {S.~A.}\ \bibnamefont
  {Miller}}, \bibinfo {author} {\bibfnamefont {D.~A.}\ \bibnamefont
  {Anderson}},\ and\ \bibinfo {author} {\bibfnamefont {G.}~\bibnamefont
  {Raithel}},\ }\href {https://doi.org/10.1088/1367-2630/18/5/053017}
  {\bibfield  {journal} {\bibinfo  {journal} {New Journal of Physics}\ }\textbf
  {\bibinfo {volume} {18}},\ \bibinfo {pages} {053017} (\bibinfo {year}
  {2016})}\BibitemShut {NoStop}%
\bibitem [{\citenamefont {Jiao}\ \emph {et~al.}(2016)\citenamefont {Jiao},
  \citenamefont {Han}, \citenamefont {Yang}, \citenamefont {Li}, \citenamefont
  {Raithel}, \citenamefont {Zhao},\ and\ \citenamefont
  {Jia}}]{29-PhysRevA.94.023832}%
  \BibitemOpen
  \bibfield  {author} {\bibinfo {author} {\bibfnamefont {Y.}~\bibnamefont
  {Jiao}}, \bibinfo {author} {\bibfnamefont {X.}~\bibnamefont {Han}}, \bibinfo
  {author} {\bibfnamefont {Z.}~\bibnamefont {Yang}}, \bibinfo {author}
  {\bibfnamefont {J.}~\bibnamefont {Li}}, \bibinfo {author} {\bibfnamefont
  {G.}~\bibnamefont {Raithel}}, \bibinfo {author} {\bibfnamefont
  {J.}~\bibnamefont {Zhao}},\ and\ \bibinfo {author} {\bibfnamefont
  {S.}~\bibnamefont {Jia}},\ }\href
  {https://doi.org/10.1103/PhysRevA.94.023832} {\bibfield  {journal} {\bibinfo
  {journal} {Phys. Rev. A}\ }\textbf {\bibinfo {volume} {94}},\ \bibinfo
  {pages} {023832} (\bibinfo {year} {2016})}\BibitemShut {NoStop}%
\bibitem [{\citenamefont {Rotunno}\ \emph {et~al.}(2023)\citenamefont
  {Rotunno}, \citenamefont {Berweger}, \citenamefont {Prajapati}, \citenamefont
  {Simons}, \citenamefont {Artusio-Glimpse}, \citenamefont {Holloway},
  \citenamefont {Jayaseelan}, \citenamefont {Potvliege},\ and\ \citenamefont
  {Adams}}]{30-10.1063/5.0162101}%
  \BibitemOpen
  \bibfield  {author} {\bibinfo {author} {\bibfnamefont {A.~P.}\ \bibnamefont
  {Rotunno}}, \bibinfo {author} {\bibfnamefont {S.}~\bibnamefont {Berweger}},
  \bibinfo {author} {\bibfnamefont {N.}~\bibnamefont {Prajapati}}, \bibinfo
  {author} {\bibfnamefont {M.~T.}\ \bibnamefont {Simons}}, \bibinfo {author}
  {\bibfnamefont {A.~B.}\ \bibnamefont {Artusio-Glimpse}}, \bibinfo {author}
  {\bibfnamefont {C.~L.}\ \bibnamefont {Holloway}}, \bibinfo {author}
  {\bibfnamefont {M.}~\bibnamefont {Jayaseelan}}, \bibinfo {author}
  {\bibfnamefont {R.~M.}\ \bibnamefont {Potvliege}},\ and\ \bibinfo {author}
  {\bibfnamefont {C.~S.}\ \bibnamefont {Adams}},\ }\href
  {https://doi.org/10.1063/5.0162101} {\bibfield  {journal} {\bibinfo
  {journal} {Journal of Applied Physics}\ }\textbf {\bibinfo {volume} {134}},\
  \bibinfo {pages} {134501} (\bibinfo {year} {2023})}\BibitemShut {NoStop}%
\bibitem [{\citenamefont {Holloway}\ \emph {et~al.}(2017)\citenamefont
  {Holloway}, \citenamefont {Simons}, \citenamefont {Gordon}, \citenamefont
  {Dienstfrey}, \citenamefont {Anderson},\ and\ \citenamefont
  {Raithel}}]{31-10.1063/1.4984201}%
  \BibitemOpen
  \bibfield  {author} {\bibinfo {author} {\bibfnamefont {C.~L.}\ \bibnamefont
  {Holloway}}, \bibinfo {author} {\bibfnamefont {M.~T.}\ \bibnamefont
  {Simons}}, \bibinfo {author} {\bibfnamefont {J.~A.}\ \bibnamefont {Gordon}},
  \bibinfo {author} {\bibfnamefont {A.}~\bibnamefont {Dienstfrey}}, \bibinfo
  {author} {\bibfnamefont {D.~A.}\ \bibnamefont {Anderson}},\ and\ \bibinfo
  {author} {\bibfnamefont {G.}~\bibnamefont {Raithel}},\ }\href
  {https://doi.org/10.1063/1.4984201} {\bibfield  {journal} {\bibinfo
  {journal} {Journal of Applied Physics}\ }\textbf {\bibinfo {volume} {121}},\
  \bibinfo {pages} {233106} (\bibinfo {year} {2017})}\BibitemShut {NoStop}%
\end{thebibliography}%

\end{document}